\documentclass[useAMS,usenatbib]{mn2e}
\usepackage{times,graphicx,float}
\usepackage{psfig}

\title[A model for 3:2 HFQPO pairs]
  {A model for 3:2 HFQPO pairs in black hole binaries based on cosmic battery}
\author[C.-Y. Huang et al.]
  {Chang-Yin Huang$^{1,2}$, Yong-Chun Ye$^2$, Ding-Xiong Wang$^2$\thanks{E-mail: dxwang@mail.hust.edu.cn} and Yang Li$^3$ \\
  $^1$School of Physics and Optoelectronic Engineering, Yangtze University, 434023, Jingzhou, China\\
  $^2$School of Physics, Huazhong University of Science and Technology, Wuhan 430074, China\\
  $^3$Department of Mathematics and Physics, Shijiazhuang Tiedao University, 050043, Shijiazhuang, China\\
 }
\date{3 December 2015}

\pagerange{\pageref{firstpage}--\pageref{lastpage}} \pubyear{0000}

\def\LaTeX{L\kern-.36em\raise.3ex\hbox{a}\kern-.15em
    T\kern-.1667em\lower.7ex\hbox{E}\kern-.125emX}

\begin{document}

\label{firstpage}

\maketitle

\begin{abstract}
   A model for 3:2 high-frequency quasi-periodic oscillations (HFQPOs) with 3:2 pairs observed in four black hole X-ray binaries (BHXBs) is proposed by invoking the epicyclic resonances with the magnetic connection (MC) between a spinning black hole (BH) with a relativistic accretion disc. It turns out that the MC can be worked out due to Poynting-Robertson cosmic battery (PRCB), and the 3:2 HFQPO pairs associated with the steep power-law states can be fitted in this model. Furthermore, the severe damping problem in the epicyclic resonance model can be overcome by transferring energy from the BH to the inner disc via the MC process for emitting X-rays with sufficient amplitude and coherence to produce the HFQPOs. In addition, we discuss the important role of the magnetic field in state transition of BHXBs.
\end{abstract}


\section{Introduction}

  It is widely accepted that accretion rate is a key parameter in governing the state transitions of black hole X-ray binaries (BHXBs). However, the main features cannot be described only by accretion rate, while magnetic fields are regarded as another key parameter in state transitions of BHXBs \citep{spru05,king12}. Unfortunately, the origin of magnetic fields in BH systems has been a puzzle in astrophysics.

  One of the most prevalent models for the origin of magnetic fields in accretion flows is the dynamo mechanism, by which the large scale ordered magnetic fields are produced from the seed magnetic fields frozen in the turbulent conducting fluid \citep{moff78,park79}. The problems with the dynamo mechanism lie in two aspects: (i) the efficiency of amplification is too low to create enough strong magnetic fields required by observations, and (ii) the uncertainty of viscosity in turbulent flow gives rise to the uncertain seed magnetic fields, and thus the uncertain large-scale magnetic fields \citep{vain92}.

  Another model for the origin of magnetic fields is the so-called Poynting-Robertson cosmic battery (PRCB) proposed by \cite{cont98}, which is based on the Poynting-Robertson drag effect on the electrons of the innermost plasma orbiting a BH or neutron star. The revisits and modifications of the PRCB mechanism and its application to astrophysics are given in a series of works (e.g., \citealt{cont06,chri08,cont09}). Recently, \cite{kyla12} apply the PRCB mechanism to the formation of magnetic fields in AGNs, and investigate whether the PRCB mechanism can also explain the formation, destruction, and variability of jets in BHXBs.

  As is well known, the high-frequency quasi-periodic oscillations (HFQPOs; 40---450 Hz) have been observed in several BHXBs \citep{RM06}. It is widely accepted that HFQPOs probably occur near the innermost stable circular orbit (ISCO), because their frequencies are well expected for matter orbiting near the ISCO for a BH of $\sim$10 solar mass. The most interesting HFQPOs is the 3:2 HFQPO pairs observed in a few BH binaries, e.g., GRO J1655$-$40 (450, 300 Hz; \citealt{remi99,stro01,remi02}), XTE J1550$-$564 (276, 184 Hz; \citealt{mill01,remi02}) and GRS 1915 + 105 (168, 113 Hz; RM06). Furthermore, the 3:2 HFQPO pair has been observed in the bright X-ray transient H1743$-$322 (240, 160 Hz; \citealt{homa05,remi06}), although the mass of its BH primary has not been measured (RM06).

  The 3:2 HFQPO pairs could be interpreted in some epicyclic resonance models \citep{AK01,abra03,kluz04,toro05}. However, there remain serious uncertainties as to whether epicyclic resonance could overcome the severe damping forces and emit X-rays with sufficient amplitude and coherence to produce the HFQPOs (e.g. see a review by \citealt{MR06}). Not long ago, \cite{huan10} applied epicyclic resonances to the magnetic connection (MC) of a BH with its surrounding relativistic accretion disc. It turns out that the 3:2 HFQPO pairs are associated with the steep power-law (SPL) states, and the above problems with the epicyclic resonance model can be overcome by transferring energy and angular momentum from a spinning BH to the inner disc in the MC process. The fittings for the 3:2 HFQPO pairs in HGWW10 are based on a model of magnetically induced disc-corona model given by \cite{gan09}. However, the origin of the MC configuration was not addressed in this work.

  Motivated by the above work, we intend to fit the 3:2 HFQPO pairs by invoking the PRCB mechanism in this paper. It turns out that the MC configuration can be created based on the electric current produced by the PRCB mechanism, and the 3:2 HFQPO pairs observed in the above sources can be well fitted associated with the corresponding spectra of the SPL states. This paper is organized as follows. In Section 2, we discuss the magnetic field configuration arising from the electric current created by the PRCB mechanism. In Section 3, we discuss the transfer of energy and angular momentum in the MC process. In Section 4, we fit the 3:2 HFQPO pairs associated with the SPL spectra of the sources by combining the MC process with epicyclic resonance model (ERM) and relativistic precession model (RPM). Finally, in Section 5, we discuss the advantages of this model over HGWW10 and the role of the magnetic field in state transitions of BHXBs.

\section{ORIGIN OF MAGNETIC FIELD AND PRCB CURRENT}

   According to the PRCB mechanism, the electric current arises from difference in the radiation-drag forces between ions and electrons, and it flows near ISCO. Thus the current density can be written as
   \begin{equation}\label{eq1}
     j_{\textup{\scriptsize PRCB}}=nev_{\textup{\scriptsize PRCB}}=ne(v_{\textup{\scriptsize i}}-v_{\textup{\scriptsize e}}),
   \end{equation}
   where $v_{\textup{\scriptsize i}}$ and $v_{\textup{\scriptsize e}}$ are the azimuthally average velocities of ion and electron, respectively, and $n$ and $e$ are the number density and electric quantity of electron, respectively. Since the ions feel a much weaker radiation-drag force than the electrons because the Thompson cross-section is inversely proportional to the square of the mass of the scatterer, i.e., $f_{\textup{\scriptsize pe}}/f_{\textup{\scriptsize pi}}=(m_{\textup{\scriptsize i}}/m_{\textup{\scriptsize e}})^2$ , where $f_{\textup{\scriptsize pi}}$ and $f_{\textup{\scriptsize pe}}$ are respectively the radiation-drag forces acting on ions and electrons, and $m_{\textup{\scriptsize i}}$ and $m_{\textup{\scriptsize e}}$ are respectively the masses of the ion and electron.

   As a simple analysis, neglecting the mass variation of electron/ion after scattering by the photon near ISCO, we have the ratio of the variation of electron's velocity to that of ion's one as follows,
   \begin{equation}\label{eq2}
     \Delta v_{\textup{\scriptsize e}}/\Delta v_{\textup{\scriptsize i}}=(m_{\textup{\scriptsize i}}/m_{\textup{\scriptsize e}})^3\simeq 6.2\times10^9\gg1.
   \end{equation}
   Thus we infer that the velocity variation of an ion after scattering with a photon can be neglected. Assuming that both an ion and an electron move initially with the Keplerian velocity $v_{\textup{\scriptsize K}}$, we can estimate the difference between the velocity of an electron and that of an ion after scattering with a photon as $\Delta v_{\textup{\scriptsize e}}=v_{\textup{\scriptsize e}}-v_{\textup{\scriptsize i}}\simeq v_{\textup{\scriptsize e}}-v_{\textup{\scriptsize K}}$, and the current density given by equation (1) can be rewritten as
      \begin{equation}\label{eq3}
    j_{\textup{\scriptsize PRCB}}=e n f_{\textup{\scriptsize pe}} \Delta t/m_{\textup{\scriptsize e}},
      \end{equation}
   where $\Delta t$ is the average time of the scattering of electrons with photons. Therefore the average radiation-drag force on the electrons can be written as
      \begin{equation}\label{eq4}
      f_{\textup{\scriptsize pe}}=\frac{F\sigma_{\textup{\scriptsize T}}}{c}\frac{v_{\textup{\scriptsize K}}}{c},
      \end{equation}
   where $F$ is the radiation flux from the relativistic accretion disc (e.g., \citealt{NT73,PT74}). In equation (4) $\sigma_{\textup{\scriptsize T}}$ is the electron Thomson cross-section, and $v_{\textup{\scriptsize K}}/c$ is the Poynting-Robertson aberration effect caused by photons ``hitting" the moving electrons on the other side at about 90$^\circ$. As a simple analysis, we ignore the geometric factor given in KCKC12 to account for light bending near the BH. Incorporating equations (3) and (4), we have the current intensity created by the PRCB mechanism as follows,
      \begin{equation}\label{eq5}
      I_{\textup{\scriptsize PRCB}}=\Delta s  j_{\textup{\scriptsize PRCB}}=\frac{e n\Delta t\Delta s}{m_{\textup{\scriptsize e}}}\frac{F \sigma_{\textup{\scriptsize T}}v_{\textup{\scriptsize K}}}{c^2},
      \end{equation}
   where $\Delta s$ is the cross-section of the current, and  $\Delta t$ is average scattering time of electrons with photons.

   How to constrain the current intensity is crucial for fitting the 3:2 HFQPO pairs observed in the above sources, because the energy transferred magnetically from a spinning BH to the inner disc depends on the strength of the MC, and thus on the strength of $ I_{\textup{\scriptsize PRCB}}$. Since the quantities $e$, $m_{\textup{\scriptsize e}}$, $\sigma_{\textup{\scriptsize T}}$  and $c$ in equation (5) are known, we can estimate the current intensity based on the following quantities, i.e.,   $\Delta s$,   $\Delta t$, $n$, $F$ and $v_{\textup{\scriptsize K}}$.

   (i) The cross-section of  $\Delta s$ is estimated as  $\Delta s=(\alpha_{\textup{\scriptsize R}}R_{\textup{\scriptsize g}})^2$, where $R_{\textup{\scriptsize g}}\equiv GM_{\textup{\scriptsize BH}}/c^2$  is the gravitational radius and $M_{\textup{\scriptsize BH}}$ is the BH mass. Thus we can obtain the values of  $\Delta s$ , such as  $\Delta s\sim10^8$cm$^2$ for the BH mass of 10 solar mass with $\alpha_{\textup{\scriptsize R}}=0.01$. For simplicity, we set  $\alpha_{\textup{\scriptsize R}}=0.01$ and regard the radius of the PRCB current   $ r_{\textup{\scriptsize PRCB}} $ as an adjustable parameter, since both $\Delta s$  and $r_{\textup{\scriptsize PRCB}} $ will affect the intensity of $ I_{\textup{\scriptsize PRCB}}$.

   (ii) The average scattering time of electrons with photons is taken as the time lag of photoelectric effect, and we have $\Delta t$  less than 10$^{-9}$ sec.

   (iii) The values of the radiation flux, $F$, the number density of the electrons, $n$, and the Kplerian velocity,  $v_{\textup{\scriptsize K}}$, can be determined in the frame of an relativistic accretion disc (see Section 3), and these values are all evaluated at radius   $r_{\textup{\scriptsize PRCB}} $.

For a relativistic accretion disc, the radiation flux $F$ varies non-monotonically with the disc radius near ISCO and reaches its maximum value $F_{\textup{\scriptsize peak}}$ at radius  $r_{\textup{\scriptsize peak}}$ \citep{NT73,PT74}. It turns out that $r_{\textup{\scriptsize PRCB}}$ varies between $r_{\textup{\scriptsize ms}} $ and $r_{\textup{\scriptsize peak}} $, being required by the fittings of the 3:2 HFQPO associated with the spectra of the SPL state, where $r_{\textup{\scriptsize ms}} $ is the radius of ISCO. Thus the PRCB current can be estimated by equation (5).

As a test of our model, we calculate the electric current by using equation (5), and the corresponding magnetic field configuration can be figured out in the frame of general relativity \citep{line79,li02}. The curves of the intensities of the current and magnetic field at ISCO versus accretion rate are shown in the left panel of Fig. 1, and magnetic configuration for  $\dot {m}=0.01$ in terms of the Eddington accretion rate $\dot {M}_{\textup{\scriptsize Edd}}=1.38\times10^{18}(M_{\textup{\scriptsize BH}} /M_\odot)$ g s$^{-1}$ are shown in the right panel of Fig. 1. The BH mass $M_{\textup{\scriptsize BH}}=10M_\odot$, spin  $a_*=0.5$ and viscosity parameter $\alpha=0.2$ are adopted in calculations. As shown in Fig. 1, two kinds of MC configurations are created by the PRCB current, i.e., the MC of the BH with the disc (MCHD) and the MC of the plunging region with the disc (MCPD) as argued in our previous work (Zhao, Wang \& Gan 2009).

The geometric units $G = c = 1$ are used in the following sections.

  \begin{figure*}
   \centering
   \includegraphics[width=5cm]{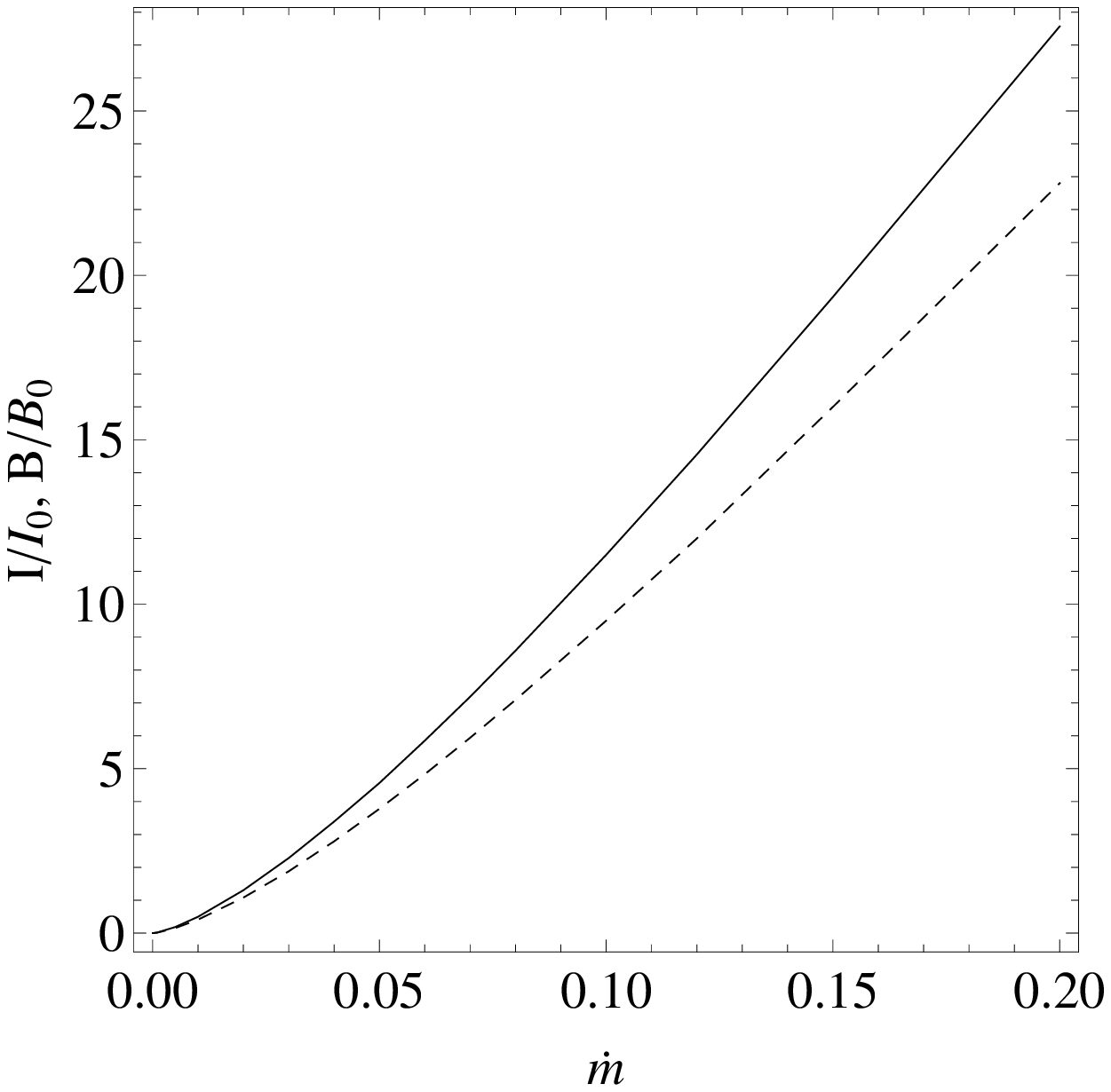}
   \hspace{0.5cm}
    \includegraphics[width=5cm]{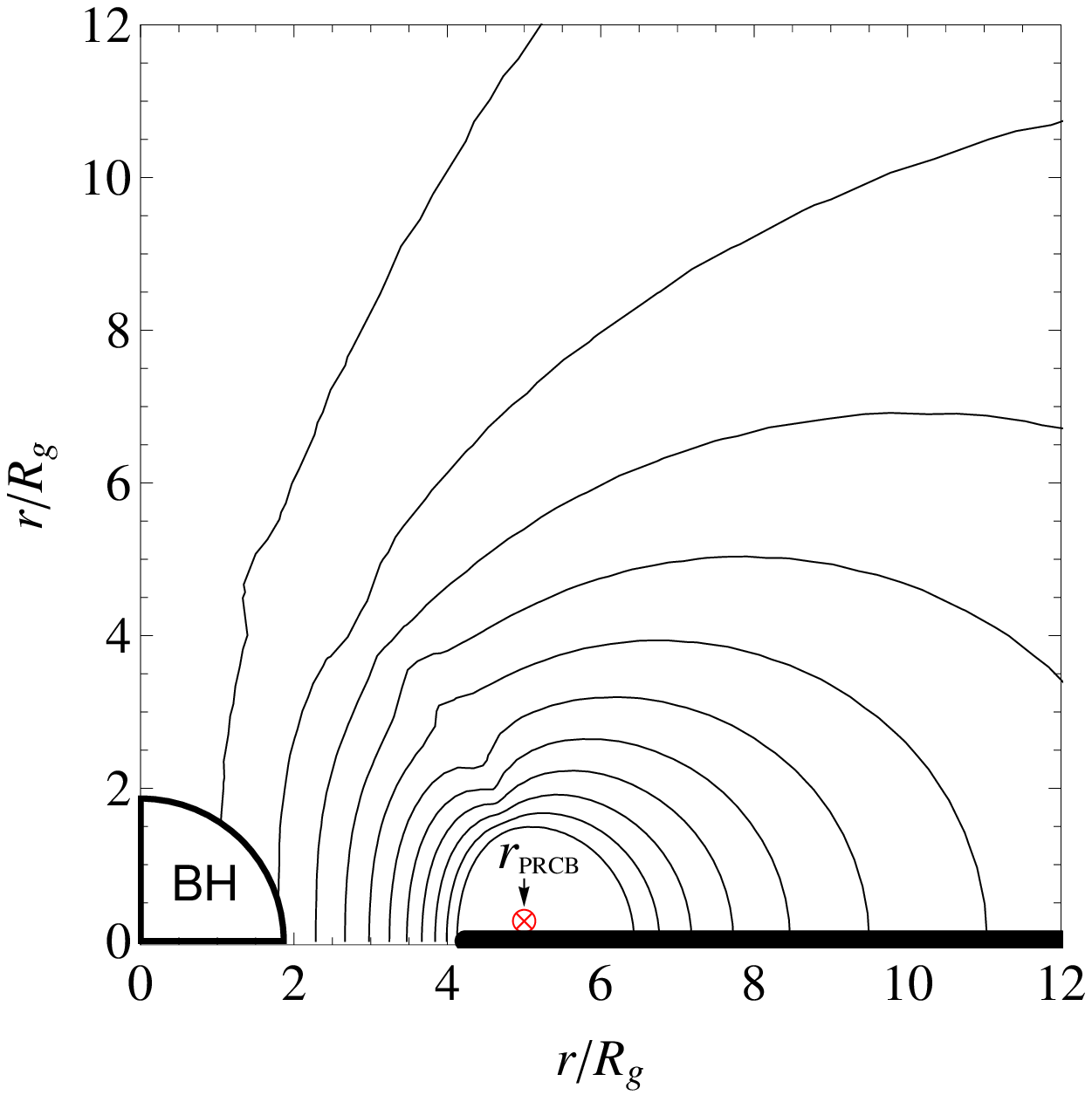}
   \caption{Left panel: curves of intensities of PRCB current (solid line) and magnetic field at ISCO (dashed line) versus accretion rate. The units of the current and magnetic field are $I_0=B_0(GM_{\textup{\scriptsize BH}}/c)$ and $B_0=10^8$ Gauss, respectively. Right panel: magnetic field configuration produced by the current at $r_{\textup{\scriptsize PRCB}} = 5R_{\textup{\scriptsize g}}$ (indicated by the symbol `$\otimes$') for $\dot{m}=0.01$. Parameters: $M_{\textup{\scriptsize BH}}=10M_\odot$, $a_*=0.5$ and $\alpha=0.2$.}
  \end{figure*}

\section{TRANSFER OF ENERGY AND ANGULAR MOMENTUM IN MC PROCESS}

Based on the conservation of magnetic flux, \cite{wang07} have the mapping relations between the polar angle $\theta$ at the BH horizon and the disc radius corresponding to MCHD and MCPD as follows,
\begin{equation}\label{eq6}
  \Psi(r,\pi/2)=\Psi(r^\prime,\pi/2), \ \ \ \ r_1<r<r_2,\ r_{\textup{\scriptsize H}}<r^\prime<r_{\textup{\scriptsize ms}},
\end{equation}
\begin{equation}\label{eq7}
  \Psi(r,\pi/2)=\Psi(r_{\textup{\scriptsize H}},\theta), \ \ \ \ r_2<r<r_{\textup{\scriptsize out}},
\end{equation}
Equations (6) and (7) correspond respectively to MCPD and MCHD, and $r_1$ and $r_2$  can be determined by relations $\Psi(r_1,\pi/2)=\Psi(r_{\textup{\scriptsize ms}},\pi/2)$  and $\Psi(r_2,\pi/2)=\Psi(r_{\textup{\scriptsize H}},\pi/2)$, respectively. The quantities $r_{\textup{\scriptsize H}}$ and $r_{\textup{\scriptsize out}}$ are the radii corresponding to the BH horizon and the outer boundary of the MCHD region, and we take $r_{\textup{\scriptsize out}}$ equal to 100 gravitational radius in this paper.

The electromotive forces in the MC process due to the rotation of the BH, the plunging region and the disc are given as follows \citep{li00,wang02},
\begin{equation}\label{eq8}
  \varepsilon_{\textup{\scriptsize H}}=\frac{\Omega_{\textup{\scriptsize H}}}{2\pi}\textup{d}\Psi, \ \ \varepsilon_{\textup{\scriptsize P}}=\frac{\Omega_{\textup{\scriptsize P}}}{2\pi}\textup{d}\Psi, \ \ \varepsilon_{\textup{\scriptsize D}}=-\frac{\Omega_{\textup{\scriptsize D}}}{2\pi}\textup{d}\Psi,
\end{equation}
where  $\textup{d}\Psi$ is the magnetic flux between the two adjacent magnetic surfaces, and  $\Omega_{\textup{\scriptsize H}}$, $\Omega_{\textup{\scriptsize D}}$  and $\Omega_{\textup{\scriptsize P}}$  are the angular velocities of the BH horizon, the disc and the plunging region, respectively \citep{MT82,wang07}:
\begin{equation}\label{eq9}
  \Omega_{\textup{\scriptsize H}}=a_*/(2r_{\textup{\scriptsize H}}),
\end{equation}
\begin{equation}\label{eq10}
  \Omega_{\textup{\scriptsize D}}=\frac{1}{M_{\textup{\scriptsize BH}}}\frac{1}{\tilde{r}^{3/2}+a_*},
\end{equation}
\begin{equation}\label{eq11}
  \Omega_{\textup{\scriptsize P}}=\frac{1}{M_{\textup{\scriptsize BH}}}\frac{(\tilde{r}-2)(L_{\textup{\scriptsize ms}}/M_{\textup{\scriptsize BH}})+2a_*E_{\textup{\scriptsize ms}}}{(\tilde{r}^3+a_*^2 \tilde{r}+2a_*^2)E_{\textup{\scriptsize ms}}-2a_*(L_{\textup{\scriptsize ms}}/M_{\textup{\scriptsize BH}})},
\end{equation}
where $\tilde{r}\equiv r/M_{\textup{\scriptsize BH}}$, and $E_{\textup{\scriptsize ms}}$ and $L_{\textup{\scriptsize ms}}$  are the specific energy and angular momentum of the accreting particles at ISCO, respectively \citep{bard72}.

The torque of the BH exerting on the disc at $r>r_2$  is
\begin{equation}\label{eq12}
  \textup{d}T_{\textup{\scriptsize HD}}=(\frac{\textup{d}\Psi}{2\pi})^2\frac{\Omega_{\textup{\scriptsize H}}-\Omega_{\textup{\scriptsize D}}}{\textup{d}Z_{\textup{\scriptsize H}}},
\end{equation}
where $\textup{d}Z_{\textup{\scriptsize H}}=2\rho_{\textup{\scriptsize H}}\textup{d}\theta/\varpi$  is the resistance at the region $(\theta,\theta+\textup{d}\theta)$ of the BH horizon, and $\rho_{\textup{\scriptsize H}}=(r_{\textup{\scriptsize H}}^2+a^2\cos^2\theta)^{1/2}$, $\varpi=(2M_{\textup{\scriptsize BH}}r_{\textup{\scriptsize H}}/\rho_{\textup{\scriptsize H}})\sin\theta$. The torque of the plunging region exerting on the disc at $r_1<r<r_2$  is
\begin{equation}\label{eq13}
  \textup{d}T_{\textup{\scriptsize PD}}=(\frac{\textup{d}\Psi}{2\pi})^2\frac{\Omega_{\textup{\scriptsize P}}-\Omega_{\textup{\scriptsize D}}}{{\rm d}Z_{\textup{\scriptsize P}}},
\end{equation}
where $\textup{d}Z_{\textup{\scriptsize P}}$ is the resistance at the region $(r^\prime,r^\prime+\textup{d}r^\prime)$ of the plunging region \citep{wang07}.

Considering the transfer of energy and angular momentum in the MC process, we have the dynamical equations of relativistic accretion disc as given by GWL09,
\begin{equation}\label{eq14}
  \frac{\textup{d}}{\textup{d}r}(\dot{M}L^\dag-g)=4\pi r(QL^\dag-H),
\end{equation}
\begin{equation}\label{eq15}
  \frac{\textup{d}}{\textup{d}r}(\dot{M}E^\dag-g\Omega_{\textup{\scriptsize D}})=4\pi r(QL^\dag-H\Omega_{\textup{\scriptsize D}}),
\end{equation}
\begin{equation}\label{eq16}
  Q\equiv Q_{\textup{\scriptsize DA}}+Q_{\textup{\scriptsize MC}}.
\end{equation}

In equations (14) and (15) the quantity $Q$ is the total energy dissipation in the disc, and $Q_{\textup{\scriptsize DA}}$  and $Q_{\textup{\scriptsize MC}}$  are the energy dissipation due to disc accretion and the MC process, respectively. The quantity $H\equiv\frac{1}{4\pi r}\frac{{\rm d}T}{{\rm d}r}$  is the flux of angular momentum transferred to the accretion disc in the MC process, and $\dot{M}$  and $g$ are accretion rate and viscous torque, respectively. Incorporating equations (14) and (15), we have
\begin{equation}\label{eq17}
  g=-\frac{\textup{d}\Omega_{\textup{\scriptsize D}}}{\textup{d}r}(E^\dag-\Omega_{\textup{\scriptsize D}}L^\dag)4\pi rQ.
\end{equation}

We take $\alpha$-prescription given by \cite{SS73}, and assume that the viscous pressure is comparable to magnetic pressure, i.e.,
\begin{equation}\label{eq18}
  P_{\textup{\scriptsize mag}}=\frac{B_{\textup{\scriptsize D}}^2}{8\pi}=\alpha P_{\textup{\scriptsize gas}},
\end{equation}
where  $B_{\textup{\scriptsize D}}$ is the tangled small-scale magnetic field in the disc. In addition, the gas pressure and radiation pressure are related by
\begin{equation}\label{eq19}
  P_{\textup{\scriptsize rad}}+P_{\textup{\scriptsize gas}}=\frac{1}{3}a_0T_{\textup{\scriptsize D}}^4+\frac{\rho k T_{\textup{\scriptsize D}}}{\mu m_{\textup{\scriptsize p}}},
\end{equation}
where $\rho$ is the mass density of the disc, $T_{\textup{\scriptsize D}}$ is the disc temperature, and  $a_0$, $k$, $m_{\textup{\scriptsize p}}$ and $\mu$ are respectively the radiation constant, the Boltzman constant, the proton mass and the mean atomic mass ($\mu$=0.615 is adopted).

  \begin{figure*}
   \centering
   \includegraphics[width=5cm]{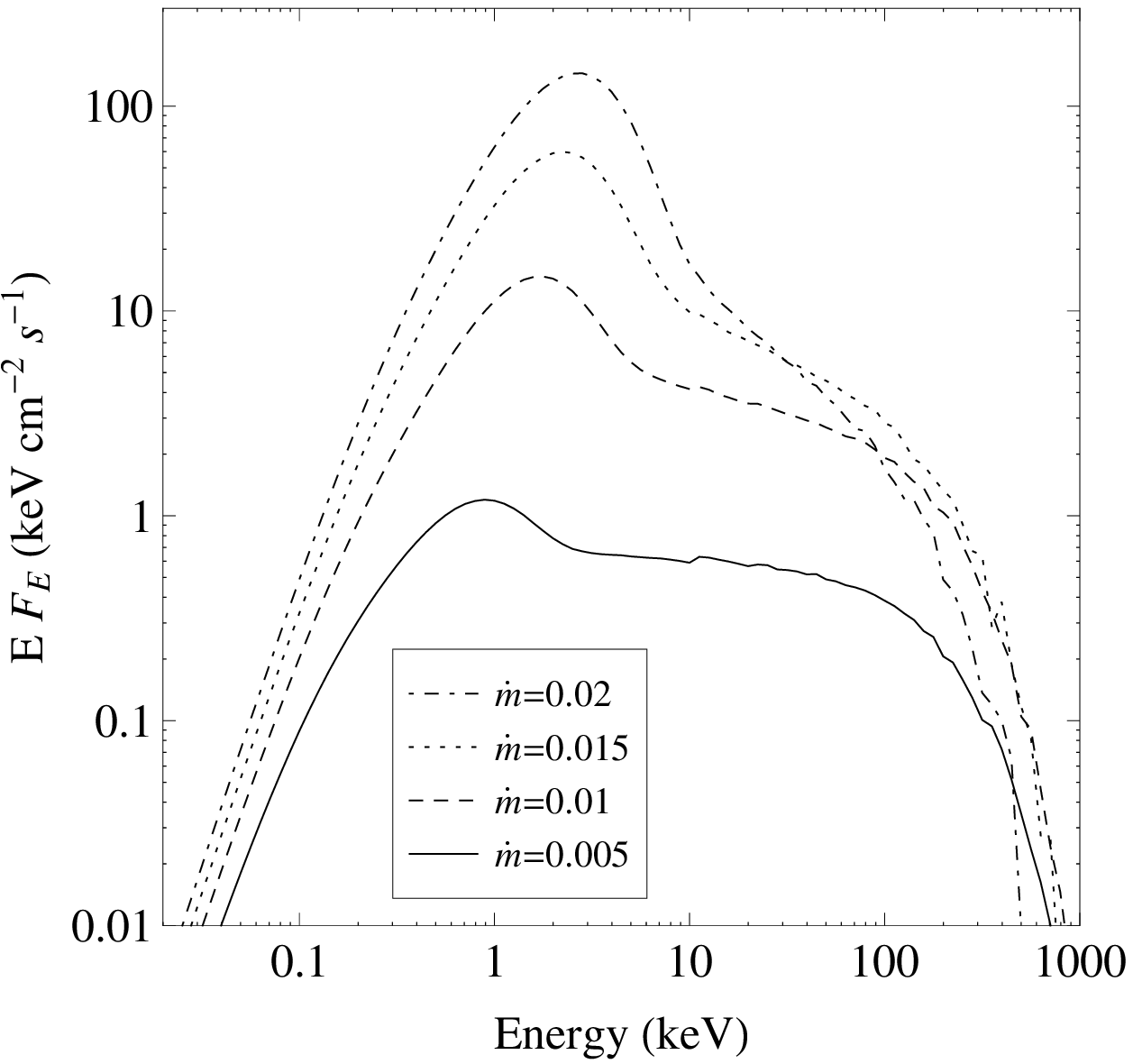}
    \hspace{0.5cm}
    \includegraphics[width=5cm]{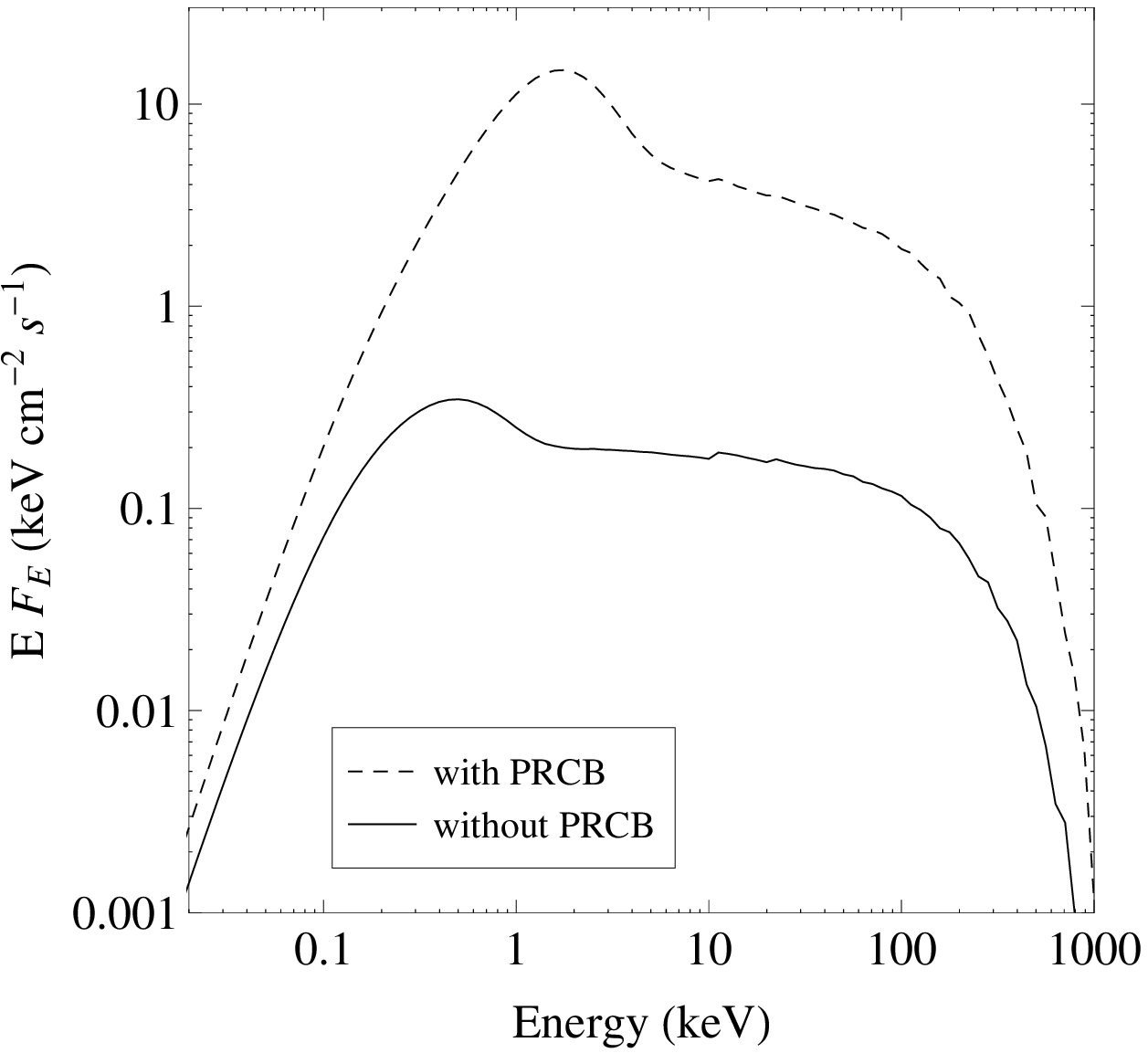}
   \caption{Emitted spectra from disc-corona system with the parameters: $M_{\textup{\scriptsize BH}}=10M_\odot$, $a_*=0.5$, $\alpha=0.2$, and  $r_{\textup{\scriptsize PRCB}}=5R_{\textup{\scriptsize g}}$. Left panel: spectra with PRCB for different accretion rates. Right panel: spectra with and without PRCB for  $\dot{m}=0.01$.}
  \end{figure*}

According to disc-corona model part of the energy dissipated in the disc is to heat corona, and the rest is released in the disc, emitting eventually as blackbody radiation and supplying seed photons for Comptonization of corona. Thus we have
\begin{equation}\label{eq20}
  Q= Q_{\textup{\scriptsize D}}+Q_{\textup{\scriptsize cor}}.
\end{equation}
In equation (20) we have $Q_{\textup{\scriptsize D}}=4\sigma T_{\textup{\scriptsize D}}^4/(3\tau_{\textup{\scriptsize D}})$ with $\sigma$ and $\tau_{\textup{\scriptsize D}}$, which are the Stefan-Boltzman constant and the optical depth, respectively. Following \cite{LMS02}, the corona is heated via magnetic reconnection of the tangled small-scale magnetic field, and we have
\begin{equation}\label{eq21}
  Q_{\textup{\scriptsize cor}}=\frac{B_{\textup{\scriptsize D}}^2}{4\pi}v_{\textup{\scriptsize A}},
\end{equation}
where $v_{\textup{\scriptsize A}}\equiv B_{\textup{\scriptsize D}}/\sqrt{4\pi\rho}$  is the local Alfven speed. On the other hand, the soft photons are scattered up by the relativistic electrons in corona, and the energy equation in corona is given as follows \citep{LMS02},
\begin{equation}\label{eq22}
  \frac{B_{\textup{\scriptsize D}}^2}{4\pi}v_{\textup{\scriptsize A}}=\frac{4kT_{\textup{\scriptsize e}}}{m_{\textup{\scriptsize e}}}\tau U_{\textup{\scriptsize rad}},
\end{equation}
where $U_{\textup{\scriptsize rad}}=a_0T_{\textup{\scriptsize D}}^4$ is the radiant energy density at vicinity of disc surface, and $\tau$ and $T_{\textup{\scriptsize e}}$ are the optical depth of the corona and temperature of the electrons in the corona, respectively.

Incorporating equations (7)---(22), we have the global solution of the disc-corona system with the MC process, and the emitted spectra of the disc-corona system can be worked out via Monte Carlo method to fit the spectra in different state observed in BHXBs (GWL09; HGWW10; \citealt{huan13,huan14}).

\section{FITTING 3:2 HFQPO PAIRS BASED ON ERM AND RPM WITH MC PROCESS}

In order to interpret the 3:2 HFQPOs pairs epicyclic resonance model (ERM, \citealt{AK01}) and relativistic precession model (RPM, \citealt{stel98,stel99,stel99b}) are invoked, in which three basic frequencies are given in GR frame as follows,
\begin{equation}\label{eq23}
  \nu_\phi=\frac{1}{2\pi}\frac{1}{M_{\textup{\scriptsize BH}}}\frac{1}{\tilde{r}^{3/2}+a_*},
\end{equation}
\begin{equation}\label{eq24}
  \nu_\theta=\nu_\phi\left(1-4a_*\tilde{r}^{-3/2}+3a_*^2\tilde{r}^{-2}\right)^{1/2},
\end{equation}
\begin{equation}\label{eq25}
  \nu_r=\nu_\phi\left(1-6\tilde{r}^{-1}+8a_*\tilde{r}^{-3/2}-3a_*^2\tilde{r}^{-2}\right)^{1/2},
\end{equation}
where  $\nu_\phi$, $\nu_\theta$  and $\nu_r$  are respectively the Keplerian frequency, the vertical and radial epicyclic frequencies. And the periastron and the nodal precession frequencies can be expressed in terms of the three basic frequencies, i.e.,
\begin{equation}\label{eq26}
  \nu_{\textup{\scriptsize per}}=\nu_\phi-\nu_r,
\end{equation}
\begin{equation}\label{eq27}
  \nu_{\textup{\scriptsize nod}}=\nu_\phi-\nu_\theta.
\end{equation}

According to ERM, resonance will occur, provided that any two of above frequencies are in small integer ratio at some disc radius. For simplicity, we interpret the observed 3:2 HFQPO pairs as two of the five frequencies satisfy the 3:2 ratio. The resonance is energized by the MC process, and thus produces the observed HFQPO pairs without the severe damping problem (HGWW10).

Following our previous work (GWL09; HGWW10), we take the optical depth and the height of the corona as $\tau=1$  and  $l=20M_{\textup{\scriptsize BH}}$, respectively, and three free parameters are involved in the disc-corona system, i.e., the BH spin  $a_*$, the accretion rate $\dot{m}$  and the viscous parameter $\alpha$.

The energy distribution between disc and corona, and thus the X-ray spectra of BHXBs are influenced by energy transferred from the spinning BH and its plunging region into the inner disc, which is based on the magnetic field arising from the PRCB mechanism. Yet another parameter, the radius of the PRCB current loop ($r_{\textup{\scriptsize PRCB}}$) is involved in the model. As an example, we fit the emitted spectra of the system with different values of   ranging from 0.005 to 0.02 as shown in the left panel of Fig. 2. The spectra become steeper, i.e., the spectral index ¦£ increases from 2.1 to 2.9 with the increasing accretion rate. And we compare the spectra with and without PRCB magnetic field for $\dot{m}=0.01$  in the right panel of Fig. 2.

Inspecting Fig.2, we find that the fraction of disc component increases due to the MC arising from the PRCB mechanism, and the stronger magnetic field gives rise to the softer spectra. The MC effects on the spectra can be found from equations (16) and (17), and the viscous torque in the inner disc is augmented. Furthermore, the temperature in the inner disc increases based on equation (18), by which the viscous torque is proportional to gas pressure, and the latter is proportional to the disc temperature. Thus the radiation pressure enhances very rapidly, being proportional to the fourth power of the disc temperature. In addition, the spectra are softer with magnetic field than those without magnetic field due to the cooling of MC given by equation (22).

  \begin{table*}
 \centering
 \begin{minipage}{105mm}
  \caption{Parameters of the BHXBs with 3:2 HFQPO pairs.}
  \begin{tabular}{@{}lcccccl@{}}
  \hline
   Source & $M_{\textup{\scriptsize BH}}/M_\odot$ & $a_*$ & {\it D} (kpc) & {\it i} ($^\circ$) & $\nu_{\textup{\scriptsize QPO}}$(Hz) & References \\
 \hline
 GRO J1655$-$40 &   6.3 & 0.7  & 3.2  & 70.2 & 300, 450 & 1, 2, 3, 4, 5, 6 \\
 XTE J1550$-$564 &  9.1 & 0.34 & 4.38 & 74.7 & 184, 276 & 1, 7, 8, 9, 10, 11 \\
 GRS 1915+105 &     14  & 0.98 & 11   & 66   & 113, 168 & 1, 12, 13, 14 \\
 H 1743$-$322 &     10$^{\textup{\scriptsize a}}$  & 0.2  & 8.5  & 75   & 160, 240 & 15, 16 \\
\hline
\end{tabular}
\newline $^{\textup{\scriptsize a}}$We take a 10 solar mass since the BH mass has not been dynamically constrained.
\newline References: (1) \cite{ozel10}; (2) \cite{shaf06}; (3) \cite{hjel95}; (4) \cite{hann00}; (5) \cite{stro01}; (6) \cite{remi99}; (7) \cite{stei11}; (8) \cite{hann09}; (9) \cite{mill01}; (10) \cite{remi02}; (11) \cite{bell12}; (12) \cite{mccl06}; (13) \cite{remi03}; (14) \cite{remi04}; (15) \cite{stei12}; (16) \cite{homa05}.
\end{minipage}
\end{table*}

Combining the PRCB mechanism with the disc-corona model, we can fit the 3:2 HFQPO pairs of the BHXBs: GRO J1655$-$40, XTE J1550$-$564, GRS 1915+105 and H1743$-$322, of which the basic parameters are listed in Table 1. As the BH mass of H1743$-$322 has not been dynamically constrained, we take $M_{\textup{\scriptsize BH}}=10M_\odot$. The fitting for the 3:2 HFQPO pairs associated with the spectra of the SPL state in these sources consists of two steps as follows.

\textbf{Step 1}: Fit the 3:2 HFQPO pairs, and five frequencies given by equations (23)---(27) are invoked. We can figure out the values of the BH spin   and the resonance radius   for the given BH mass and HFQPO pair listed in Table 1. It turns out that these HFQPO pairs can be well fitted, corresponding to the same resonance radius for each type of resonance in each source. There are three types of resonance for each source, and each type of resonance corresponds to one value of BH spin, and we obtain three values of BH spin for the fitting of each source as shown in Table 2.

\textbf{Step 2}: We choose the spin most close to that determined by continuum fitting method (Table 1) from the three values determined by three types of resonance for each source. And then we fit the spectrum by using the Monte Carlo method, and compare it with the observed one for each source by adjusting the accretion rate $\dot{m}$, the viscous parameter $\alpha$ and the radius of PRCB current $r_{\textup{\scriptsize PRCB}}$, and the galactic hydrogen absorption is considered in the fittings. The fitting parameters are listed in Table 3, and the associated spectra of the SPL states of these BHXBs are shown in Fig. 3, in which the total spectrum and its disc component and power law component from the corona are given.

As shown in Table 3 and Fig. 3, the spectra are fitted in accordance with the observed SPL spectra for the accretion rate of a few percent of the Eddington rate with $\alpha\sim0.2$. For XTE J1550$-$564, the values of $r_{\textup{\scriptsize PRCB}}$  and $\alpha$  are larger than those of other three sources. This is because XTE J1550$-$564 was experiencing a flare (September 19, 1998) with a very high luminosity, so a strong magnetic field is required for fitting the spectrum.

  \begin{table*}
 \centering
 \begin{minipage}{130mm}
  \caption{ Resonance frequencies and the 3:2 HFQPO pairs of BHXBs.}
  \begin{tabular}{@{}lcccccccc@{}}
  \hline
   Source & Type of resonance & $a_*$ & $r_{\textup{\scriptsize res}}$($R_{\textup{\scriptsize g}}$) & $\nu_\phi$(Hz) & $\nu_\theta$(Hz) & $\nu_r$(Hz) & $\nu_{\textup{\scriptsize per}}$(Hz) & $\nu_{\textup{\scriptsize nod}}$(Hz) \\
 \hline
 GRO J1655$-$40 &   $\nu_\theta/\nu_r=3/2$ & 0.97  & 4.2  & 534 & 450 & 300 & 234 & 84 \\
                &   $\nu_\theta/\nu_{\textup{\scriptsize per}}=3/2$ & \textbf{0.7}  & \textbf{4.4}  & \textbf{511} & \textbf{450} & \textbf{211} & \textbf{300} & \textbf{61} \\
                &   $\nu_\phi/\nu_{\textup{\scriptsize per}}=3/2$ & 0.49  & 4.9  & 450 & 415 & 150 & 300 & 35 \\
 \hline
 XTE J1550$-$564 &   $\nu_\theta/\nu_r=3/2$ & 0.93  & 4.7  & 318 & 276 & 184 & 134 & 42 \\
                 &   $\nu_\theta/\nu_{\textup{\scriptsize per}}=3/2$ & 0.53  & 5.0  & 300 & 276 & 116 & 184 & 24 \\
                 &   $\nu_\phi/\nu_{\textup{\scriptsize per}}=3/2$ & \textbf{0.38}  & \textbf{5.4}  & \textbf{276} & \textbf{261} & \textbf{92} & \textbf{184} & \textbf{15} \\
\hline
 GRS 1915+105    &   $\nu_\theta/\nu_r=3/2$ & \textbf{0.9 } & \textbf{5.0}  & \textbf{191} & \textbf{168} &
                 \textbf{112} & \textbf{79} & \textbf{23} \\
                 &   $\nu_\theta/\nu_{\textup{\scriptsize per}}=3/2$ & 0.43  & 5.4  & 179 & 168 & 67 & 112 & 11 \\
                 &   $\nu_\phi/\nu_{\textup{\scriptsize per}}=3/2$ & 0.3  & 5.7  & 168 & 161 & 56 & 112 & 7 \\
 \hline
 H 1743$-$322    &   $\nu_\theta/\nu_r=3/2$ & 0.9  & 4.9  & 274 & 240 & 160 & 114 & 34 \\
                 &   $\nu_\theta/\nu_{\textup{\scriptsize per}}=3/2$ & \textbf{0.46}  & \textbf{5.3}  & \textbf{257} & \textbf{240} & \textbf{97} & \textbf{160} & \textbf{17} \\
                 &   $\nu_\phi/\nu_{\textup{\scriptsize per}}=3/2$ & 0.33  & 5.6  & 240 & 229 & 80 & 160 & 11 \\
\hline
\end{tabular}
\end{minipage}
\end{table*}

  \begin{table*}
 \centering
 \begin{minipage}{100mm}
  \caption{Parameters for fitting energy spectra of the BHXBs with the 3:2 HFQPO pairs.}
  \begin{tabular}{@{}lccccc@{}}
  \hline
   Source  & $a_*$ & $r_{\textup{\scriptsize PRCB}}$($R_{\textup{\scriptsize g}}$) & $\dot{m}$ & $\alpha$ & $N_{\textup{\scriptsize H}}$($10^{22}$cm$^{-2}$) \\
 \hline
 GRO J1655$-$40 &  0.7 &  3.5  & 0.015 & 0.18 & 0.89$^a$ \\
 XTE J1550$-$564 &  0.38 &  5.0  & 0.025 & 0.24 & 0.32$^b$ \\
 GRS 1915+105 &  0.9 &  2.4  & 0.055 & 0.18 & 5$^c$ \\
 H 1743$-$322 &  0.46 &  4.6  & 0.024 & 0.18 & 2.3$^d$ \\
\hline
\end{tabular}
\newline $^a$\cite{shap07}; $^{b}$\cite{toms01}; $^{c}$\cite{lee02}; $^{d}$\cite{mill06}
\end{minipage}
\end{table*}

 \begin{figure*}
   \centering
   \includegraphics[width=5cm]{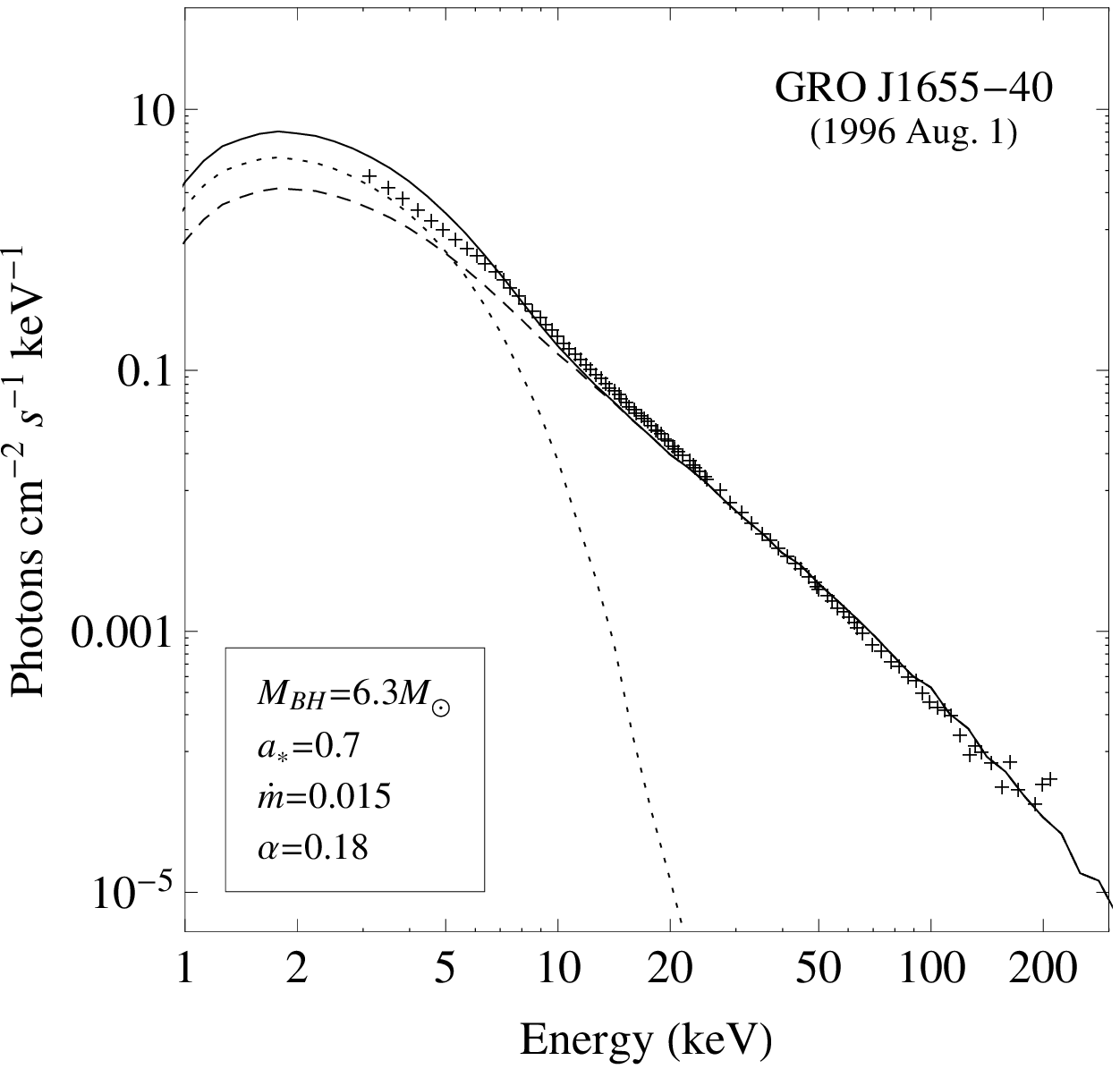}\hspace{0.5cm}\includegraphics[width=5cm]{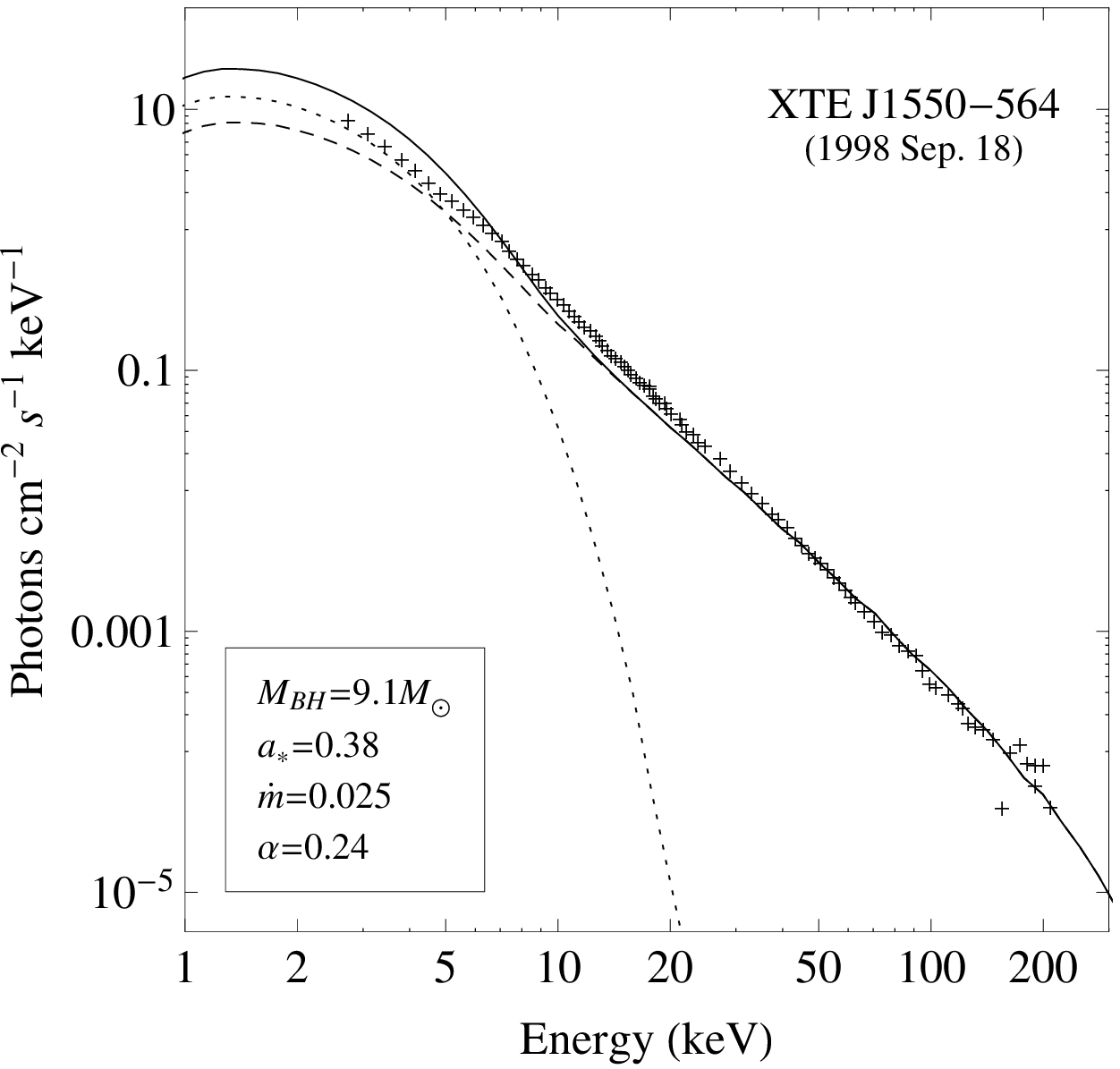}\\
   \includegraphics[width=5cm]{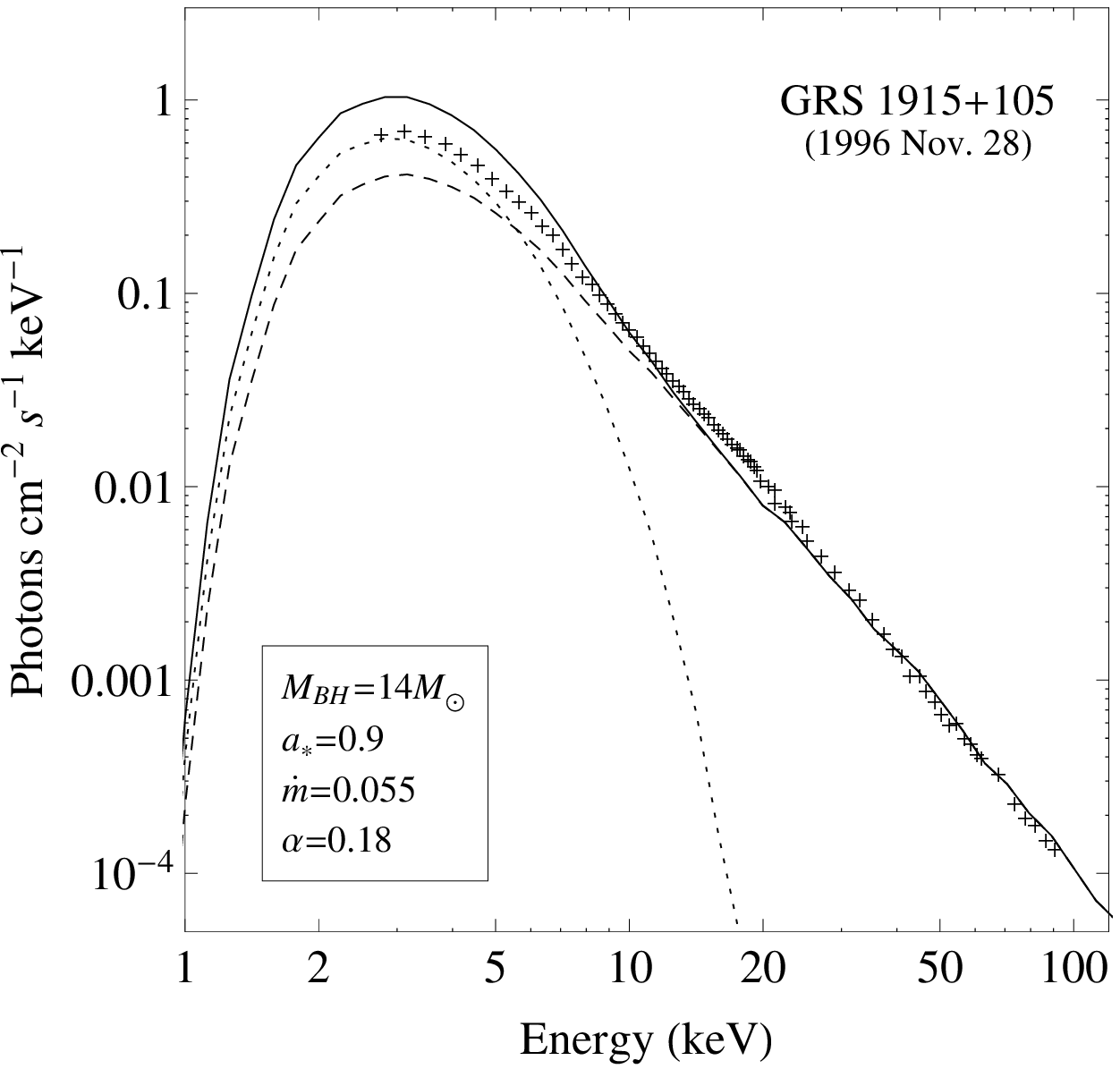}\hspace{0.5cm}\includegraphics[width=5cm]{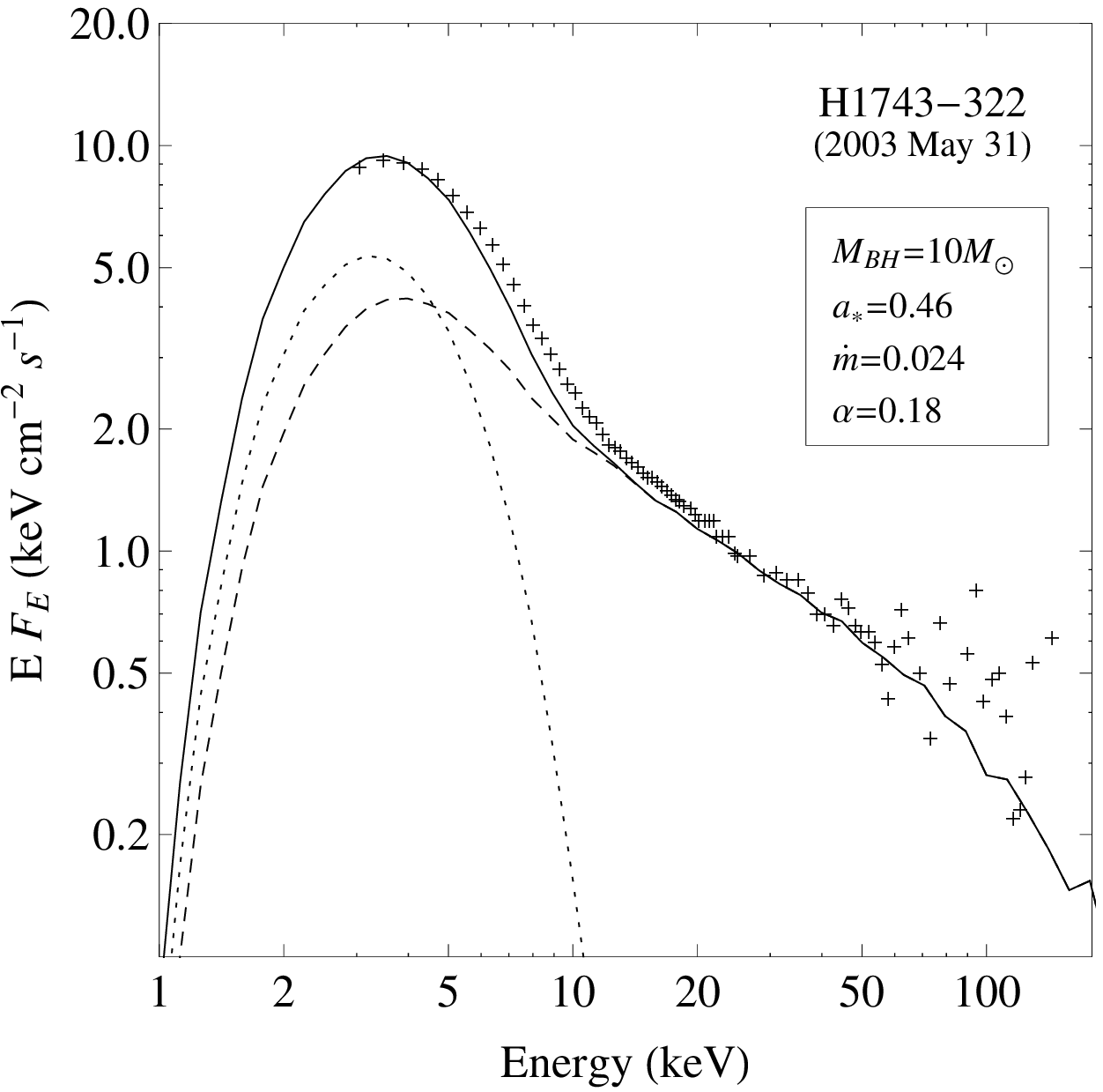}\\
   \caption{The spectra of the SPL state of the four BHXBs, where solid, dotted and dashed lines represent the total spectra, the disc component and the power-law component, respectively. The upper left, upper right, lower left and lower right panels correspond to GRO J1655$-$40, XTE J1550$-$564, GRS 1915+105 and H1743$-$322, respectively. The symbol ``+" represents the observation data adopted from \citet{remi04} for GRO J1655$-$40 and XTE J1550$-$564, and it represents the observation data adopted from \citet{MR06} and \citet{mill06} for GRS 1915+105 and H1743$-$322, respectively.}
  \end{figure*}

\section{DISCUSSION}

In this paper, we fit the 3:2 HFQPO pairs associated with the SPL states observed in the four BHXBs. The fittings are based on ERM and RPM with the large-scale magnetic fields arising from the PRCB mechanism. It turns out that the energy can be transferred from a spinning BH to the inner disc via the MC process, and the severe damping in ERM and RPM could be avoided. In addition, the spectra of the SPL states can be fitted by invoking the MC with the disc-corona model. 

Not long ago, \cite{toro12} pointed out that the relation between the frequencies of upper and lower HFQPO peaks itself, and mass and spin of the sources cannot fully be determined without other independent data. This drawback can be avoided in our model. As shown in section 4, steep power law emitted spectra are required in fitting the HFQPO pairs, which is used as an independent constraint to the BH mass and spin. Thus we can fit the HFQPO pairs associated with the spectra in SPL state of each BHXB, and the BH mass and spin are in agreement with the observations. 

\cite{baka10} considered magnetic-field-induced non-geodesic corrections to charged test particles orbiting a non-rotating neutron star. It turns out that magnetic field effects are important in the fitting of HFQPOs data for some low-mass x-ray binaries (LMXB). In addition, it is shown that the presence of magnetic interaction can significantly improve fitting of the HFQPOs data for some LMXB in the relativistic precession model \citep{baka12}. 

The situation seems somehow different for fitting HFQPO pairs of BHXBs based on PRCB effect in the following aspects. (i) Compared to neutron stars, the origin of the magnetic fields in BHXBs remains elusive, and it assumed that magnetic field could be carried by accreting plasma from a companion or produced by electric current flowing in the accretion disc due to e.g., PRCB effect. (ii) Unlike a neutron star, the magnetic fields of a BH is not intrinsic and is not so strong as a neutron star, since they must be created and maintained by its surrounding accretion disc. Thus we think that the magnetic effects on HFQPOs could not be very important as those for neutron stars. 

Now we give a rough approximate estimation on the PRCB effect in fitting HFQPOs pairs as follows. As shown in Figure 1 and Tables 2--3, the magnetic field lines tread to disc vertically, and the resonance radii $r_{\textup{\scriptsize res}}$ (indicated in boldface in Table 2) are all located outside the PRCB current. Thus, in the case in Figure 1, the Lorentz force exerted by the PRCB magnetic field makes the electrons move outwards, and makes the ions move inwards. Considering that the PRCB effect results in the average Keplerian velocity of the electrons less than that of the ions, we infer that the resultant force makes the plasma move inwards from the resonance radii $r_{\textup{\scriptsize res}}$. Based on equations (23)--(25) we infer that the PRCB effect could give rise to greater toroidal and vertical resonance frequencies (i.e.,  $\nu_\phi$ and $\nu_\theta$) and less radial resonance frequency $\nu_r$. However, it is easy to check that the resultant Lorentz force on the plasmas is much less than the gravity on them, and the ratio of the former to the latter is $\sim10^{-9}$. Thus we can neglect the PRCB effect on the HFQPO frequencies. 

Comparing with HGWW10, we have two advantages in this model as follows.

(i) The origin of the large-scale magnetic fields due to PRCB current is more reasonable than those generated from the small-scale magnetic field in dynamo process. The magnetic configuration was not accurately considered in HGWW10: the BH horizon and the inner disc are simply assumed to be connected by the closed magnetic field lines, and its strength is estimated to scale the height of the disc, and constrained by the balance between the ram pressure in the inner disc and the magnetic pressure on the BH horizon. While in this paper, we calculate the magnetic configuration and strength accurately by the PRCB mechanism. It turns out that both the horizon and the plunging region are connected to the inner disc, i.e., both magnetic connection (MCHD and MCPD) are taken into account in this model.

(ii) The model is more suitable to interpret the transition from low/hard state to the SPL state. As shown in equation (5) the PRCB current is closely related to the radiation flux, and the latter is closely related to the accretion rate. Thus we infer the MC has a positive correlation with the accretion rate, and a strong PRCB current gives rise to a strong MC process in the SPL state, in which the accretion rate is very high \citep{MR06}. While in low/hard state the PRCB current is not strong enough to produce a strong PRCB current due to the accretion rate less than $\sim0.01\dot{M}_{\textup{\scriptsize Edd}}$. This scenario could be naturally interpreted in the context of the truncated disc model (e.g., \citealt{done07}). During the hard-to-soft transition, in which the truncated radius between the inner ADAF and the outer thin disc decreases monotonously with the accretion rate and the X-ray luminosity. Therefore, a very strong radiation flux and PRCB current, and thus a very strong stronger magnetic field attain in SPL state. These results arise from a high accretion rate, resulting in a softer spectrum and a steeper power law component as shown in Fig. 2.

Furthermore, we notice that the BH spins required by the fittings of the 3:2 HFQPO pairs via the PRCB mechanism coincide very well with those measured by the continuum fitting method as shown in Table 3, being exactly the BH spins required by the BZ process for powering the relativistic, episodic jets as argued by \cite{NM12} and \cite{MN13}.

Relativistic transient jets are usually observed in the hard-to-soft transition (\citealt{fend04}; RM06), and a correlation between jet power and BH spin was discovered for five BHXBs, which means that the transient jets are powered by extracting rotational energy of BH probably via the BZ process \citep{bz77,NM12,MN13}. It is noticed that both the 3:2 HFQPO pairs and the relativistic transient jets are observed in the hard-to-soft state transitions of the four sources, GRO J1655$-$40, XTE J1550$-$564, GRS 1915+105 and H1743$-$322. This coincidence implies an intrinsic connection related to these phenomena. Considering that the transient jets are powered by the BZ process and the 3:2 HFQPOs pairs are input energy by the MC process, we think the intrinsic connection lies in the large-scale magnetic fields, which are both related closely to the extraction of energy from a spinning BH.

As a summary, we intend to emphasize the following points: (i) magnetic fields do play a very important role in state transitions of BHXBs, which could be regarded as the second parameter besides accretion rate as suggested by some authors \citep{spru05,mill12,king12,siko13,dext14,ye15}; (ii) QPO method and continuum method might not contradict in measuring BH spins, at least, for the above four sources. Further investigations and more observations are needed to test the consistence of these methods in measuring BH spins.

\section*{Acknowledgments}

This work is supported by the National Basic Research Program of China (2009CB824800) and the National Natural Science Foundation of China (grants 11173011, 11403003 and U1431101). We are very grateful to the anonymous referee for his (her) very instructive suggestion for improving our work.

\label{lastpage}


\begin{thebibliography}{99}


\bibitem[\protect\citeauthoryear{Abramowicz \& Kl\'{u}zniak}{2001}]{AK01} Abramowicz M. A., Kl$\acute{\textup{u}}$zniak W., 2001, A\&A, 374, L19

\bibitem[\protect\citeauthoryear{Abramowicz et al.}{2003}]{abra03} Abramowicz M. A., Karas V., Kl\'{u}zniak W., Lee W. H., Rebusco P., 2003, PASJ, 55, 467

\bibitem[\protect\citeauthoryear{Bakala et al.}{2010}]{baka10} Bakala P., \v{S}r\'{a}mkov\'{a} E., Stuchl\'{\i}k Z., T\"{o}r\"{o}k G., 2010, Class. Quantum Grav., 27, 045001

\bibitem[\protect\citeauthoryear{Bakala et al.}{2012}]{baka12} Bakala P., Urbanec M., \v{S}r\'{a}mkov\'{a} E., Stuchl\'{\i}k Z., T\"{o}r\"{o}k G., 2012, Class. Quantum Grav., 29, 065012

\bibitem[\protect\citeauthoryear{Bardeen, Press \& Teukolsky}{1972}]{bard72} Bardeen J. M., Press W. H., Teukolsky S. A., 1972, ApJ, 178, 347

\bibitem[\protect\citeauthoryear{Belloni, Sanna \& M\'{e}ndez}{2012}]{bell12} Belloni T. M., Sanna A., M\'{e}ndez M., 2012, MNRAS, 426, 1701

\bibitem[\protect\citeauthoryear{Blandford \& Znajek}{1977}]{bz77} Blandford R. D., Znajek R. L., 1977, MNRAS, 179, 433

\bibitem[\protect\citeauthoryear{Christodoulou, Contopoulos \& Kazanas}{2008}]{chri08} Christodoulou D. M., Contopoulos I., Kazanas D., 2008, ApJ, 674, 388

\bibitem[\protect\citeauthoryear{Contopoulos \& Kazanas}{1998}]{cont98} Contopoulos I., Kazanas D., 1998, ApJ, 508, 859

\bibitem[\protect\citeauthoryear{Contopoulos, Kazanas \& Christodoulou}{2006}]{cont06} Contopoulos I., Kazanas D., Christodoulou D. M., 2006, ApJ, 652, 1451

\bibitem[\protect\citeauthoryear{Contopoulos et al.}{2009}]{cont09}  Contopoulos I., Christodoulou D. M., Kazanas D., Gabuzda D. C., 2009, ApJ, 702, L148

\bibitem[\protect\citeauthoryear{Dexter et al.}{2014}]{dext14} Dexter J., McKinney J. C., Markoff S., Tchekhovskoy A., 2014, MNRAS, 440, 2185

\bibitem[\protect\citeauthoryear{Done, Gierli\'{n}ski \& Kubota}{2007}]{done07} Done C., Gierli\'{n}ski M., Kubota A., 2007, A\&A Rev., 15, 1

\bibitem[\protect\citeauthoryear{Fender, Belloni \& Gallo}{2004}]{fend04} Fender R. P., Belloni T. M., Gallo E., 2004, MNRAS, 355, 1105

\bibitem[\protect\citeauthoryear{Gan, Wang \& Lei}{2009, hereafter GWL09}]{gan09} Gan Z.-M., Wang D.-X., Lei W.-H., 2009, MNRAS, 394, 2310 (GWL09)

\bibitem[\protect\citeauthoryear{Hannikainen et al.}{2000}]{hann00} Hannikainen D. C., Hunstead R. W., Campbell-Wilson D., et al. 2000, ApJ, 540, 521

\bibitem[\protect\citeauthoryear{Hannikainen et al.}{2009}]{hann09}  Hannikainen D. C. et al., 2009, MNRAS, 397, 569

\bibitem[\protect\citeauthoryear{Hjellming \& Rupen}{1995}]{hjel95} Hjellming R. M., Rupen M. P., 1995, Nat, 375, 464

\bibitem[\protect\citeauthoryear{Homan et al.}{2005}]{homa05} Homan J., Miller J. M., Wijnands R., van der Klis M., Belloni T., Steeghs D., Lewin W. H. G., 2005, ApJ, 623, 383

\bibitem[\protect\citeauthoryear{Huang et al.}{2010, hereafter HGWW10}]{huan10} Huang C.-Y., Gan Z.-M., Wang J.-Z., Wang D.-X., 2010, MNRAS, 403, 1978 (HGWW10)

\bibitem[\protect\citeauthoryear{Huang et al.}{2013}]{huan13} Huang C.-Y., Wang D.-X., Wang J.-Z., Wang Z.-Y., 2013, RAA, 13, 705

\bibitem[\protect\citeauthoryear{Huang, Gong \& Wang}{2014}]{huan14} Huang C.-Y., Gong X.-L., Wang D.-X., 2014, CPL, 31, 129701

\bibitem[\protect\citeauthoryear{King et al.}{2012}]{king12}  King A. L., Miller J. M., Raymond J., et al., 2012, ApJ, 746, L20

\bibitem[\protect\citeauthoryear{Kl$\acute{\textup{u}}$zniak, Abramowicz \& Lee}{2004}]{kluz04} Kl$\acute{\textup{u}}$zniak W., Abramowicz M. A., Lee W., 2004, in Kaaret P., Swank J. H., Lamb F. K., eds, AIP Conf. Ser. Vol. 714, X-Ray Timing 2003: Rossie and Beyond. Am. Inst. Phys., New York,p. 379

\bibitem[\protect\citeauthoryear{Kylafis et al.}{2012, hereafter KCKC12}]{kyla12} Kylafis N. D., Contopoulos I., Kazanas D., Christodoulou D. M., 2012, A\&A, 538, A5 (KCKC12)

\bibitem[\protect\citeauthoryear{Lee et al.}{2002}]{lee02}  Lee J. C., Reynolds C. S., Remillard R., et al., 2002, ApJ, 567, 1102

\bibitem[\protect\citeauthoryear{Linet}{1979}]{line79} Linet B., 1979, J. Phys. A, 12, 839

\bibitem[\protect\citeauthoryear{Li}{2000}]{li00} Li L.-X., 2000, ApJ, 533, L115

\bibitem[\protect\citeauthoryear{Li}{2002}]{li02} Li L.-X., 2002, Phys. Rev. D, 65 084047

\bibitem[\protect\citeauthoryear{Liu, Mineshige \& Shibata}{2002}]{LMS02} Liu B.-F., Mineshige S., Shibata K., 2002, ApJ, 572, L173

\bibitem[\protect\citeauthoryear{MacDonald \& Thorne}{1982}]{MT82} MacDonald D., Thorne K. S., 1982, MNRAS, 198, 345

\bibitem[\protect\citeauthoryear{McClintock \& Remillard}{2006}]{MR06} McClintock J. E., Remillard R. A., 2006, in Lewin, van der Klis, eds, Compact Stellar X-ray Sources. Cambridge Univ. Press, Cambridge, p. 157

\bibitem[\protect\citeauthoryear{McClintock et al.}{2006}]{mccl06} McClintock J. E., Shafee R., Narayan R., et al., 2006, ApJ, 652, 518

\bibitem[\protect\citeauthoryear{McClintock \& Narayan}{2013}]{MN13} McClintock J. E., Narayan R., 2013, ApJ, 762, 104

\bibitem[\protect\citeauthoryear{Miller et al.}{2001}]{mill01} Miller J. M., Wijnands R., Homan J., et al., 2001, ApJ, 563, 928

\bibitem[\protect\citeauthoryear{Miller et al.}{2006}]{mill06} Miller J. M., Raymond J., Homan J., et al., 2006, ApJ, 646, 394

\bibitem[\protect\citeauthoryear{Miller et al.}{2012}]{mill12} Miller J. M., Reynolds C. S., Fabian A. C., et al., 2012, ApJ, 759, L6

\bibitem[\protect\citeauthoryear{Moffatt}{1978}]{moff78} Moffatt H. K., 1978, Magnetic Field Generation in Electrically Conducting Fluids (Cambridge: Cambridge Univ. Press)

\bibitem[\protect\citeauthoryear{Narayan \& McClintock}{2012}]{NM12} Narayan R., McClintock J. E., 2012, MNRAS, 419, L69

\bibitem[\protect\citeauthoryear{Novikov \& Thorne}{1973}]{NT73}  Novikov I. D., Thorne K. S., 1973, in Black Holes, ed. C. DeWitt \& B. S. DeWitt (New York: Gordon \& Breach), 343

\bibitem[\protect\citeauthoryear{\"{O}zel et al.}{2010}]{ozel10} \"{O}zel F., Psaltis D., Narayan R., McClintock J. E., 2010, ApJ, 725, 1918

\bibitem[\protect\citeauthoryear{Page \& Thorne}{1974}]{PT74} Page D. N., Thorne K. S., ApJ, 1974, 191, 499

\bibitem[\protect\citeauthoryear{Parker}{1979}]{park79} Parker E. N., 1979, Cosmical Magnetic Fields (Oxford: Clarendon)

\bibitem[\protect\citeauthoryear{Remillard et al.}{1999}]{remi99} Remillard R. A., Morgan E. H., McClingtock J. E., et al., 1999, ApJ, 522, 397

\bibitem[\protect\citeauthoryear{Remillard et al.}{2002}]{remi02} Remillard R. A., Muno M. P., McClintock J. E., et al., 2002, ApJ, 580, 1030

\bibitem[\protect\citeauthoryear{Remillard et al.}{2003}]{remi03} Remillard R. A., Muno M. P., McClintock J. E., Orosz
    J. A., 2003, BAAS, 35, 648

\bibitem[\protect\citeauthoryear{Remillard}{2004}]{remi04} Remillard R. A., 2004, AIPC, 714, 13

\bibitem[\protect\citeauthoryear{Remillard \& McClintock}{2006, hereafter RM06}]{RM06} Remillard R. A., McClintock J. E., 2006, ARA\&A, 44, 49 (RM06)

\bibitem[\protect\citeauthoryear{Remillard et al.}{2006}]{remi06} Remillard R. A., McClintock J. E., Orosz J. A., Levine A. M., 2006, ApJ, 637, 1002

\bibitem[\protect\citeauthoryear{Shafee et al.}{2006}]{shaf06} Shafee R., McClintock J. E., Narayan R., Davis S. W., Li L.-X., Remillard R. A., 2006, ApJ, 636, L113

\bibitem[\protect\citeauthoryear{Shakura \& Sunyaev}{1973}]{SS73} Shakura N. I., Sunyaev R. A., 1973, A\&A, 24, 337


\bibitem[\protect\citeauthoryear{Shaposhnikov et al.}{2007}]{shap07} Shaposhnikov N., Swank J., Shrader C. R., et al., 2007, ApJ, 655, 434

\bibitem[\protect\citeauthoryear{Sikora \& Begelman}{2013}]{siko13} Sikora M., Begelman M. C., 2013, ApJ, 764, L24


\bibitem[\protect\citeauthoryear{Spruit \& Uzdensky}{2005}]{spru05} Spruit H. C., Uzdensky D. A., 2005, ApJ, 629, 960

\bibitem[\protect\citeauthoryear{Steiner et al.}{2011}]{stei11} Steiner J. F. et al., 2011, MNRAS, 416, 941

\bibitem[\protect\citeauthoryear{Steiner, McClintock \& Reid}{2012}]{stei12} Steiner J. F., McClintock J. E., Reid M. J., 2012, ApJ, 745, L7

\bibitem[\protect\citeauthoryear{Stella \& Vietri}{1998}]{stel98} Stella L., Vietri M., 1998, ApJ, 492, L59+

\bibitem[\protect\citeauthoryear{Stella \& Vietri}{1999}]{stel99} Stella L., Vietri M., 1999, Physical Review Letters, 82, 17

\bibitem[\protect\citeauthoryear{Stella, Vietri \& Morsink}{1999}]{stel99b} Stella L., Vietri M., Morsink S. M., 1999, ApJ, 524, L63

\bibitem[\protect\citeauthoryear{Strohmayer}{2001}]{stro01} Strohmayer T. E., 2001 ApJ, 552, L49

\bibitem[\protect\citeauthoryear{Tomsick}{2001}]{toms01} Tomsick J. A., Corbel S., Kaaret P., 2001, ApJ, 563, 229

\bibitem[\protect\citeauthoryear{T\"{o}r\"{o}k et al.}{2005}]{toro05} T\"{o}r\"{o}k G., Abramowicz M. A., Kl\'{u}zniak W., Stuchl\'{\i}k Z., 2005, A\&A, 436, 1

\bibitem[\protect\citeauthoryear{T\"{o}r\"{o}k et al.}{2012}]{toro12} T\"{o}r\"{o}k G., Bakala P., \v{S}r\'{a}mkov\'{a} E., Stuchl\'{\i}k Z., Urbanec M., Goluchov\'{a} K., 2012, ApJ, 760, 138

\bibitem[\protect\citeauthoryear{Vainshtein \& Cattaneo}{1992}]{vain92} Vainshtein S. I., Cattaneo F., 1992, ApJ, 393, 165

\bibitem[\protect\citeauthoryear{Wang, Xiao and Lei}{2002}]{wang02} Wang D.-X., Xiao K., Lei W.-H., 2002, MNRAS, 335, 655

\bibitem[\protect\citeauthoryear{Wang et al.}{2007}]{wang07} Wang D.-X., Ye Y.-C., Li Y, Liu D.-M., 2007, MNRAS, 374, 647

\bibitem[\protect\citeauthoryear{Ye et al.}{2015}]{ye15} Ye Y.-C., Wang D.-X., Huang C.-Y., Cao X.-F., 2015, RAA, accepted, arXiv: 1508.07108

\bibitem[\protect\citeauthoryear{Zhao, Wang & Gan}{2009}]{zhao09} Zhao C.-X.,Wang D.-X., Gan, Z.-M., 2009, MNRAS, 398, 1886


\end{thebibliography}
\end{document}